\documentclass[twocolumn]{jpsj3}
\usepackage{txfonts}
\usepackage{graphicx}% Include figure files
\usepackage{dcolumn}% Align table columns on decimal point
\usepackage{color}
\definecolor{green}{rgb}{0.0, 0.5, 0.0} 
\usepackage{amssymb}
\usepackage{braket}
\usepackage{bm}

\title{Finite Temperature Properties of Geometrically Charge Frustrated Systems}
\author{Kazuyoshi \textsc{Yoshimi}$^1$, Makoto \textsc{Naka}$^2$,  and Hitoshi \textsc{Seo}$^{3, 4}$}

\inst{$^1$Institute for Solid State Physics, University of Tokyo, Chiba 277-8581\\
$^2$ Waseda Institute for Advanced Study, Waseda University, Tokyo 169-8555\\
$^3$ Condensed Matter Theory Laboratory, RIKEN, Saitama 351-0198\\
$^4$ RIKEN Center for Emergent Matter Science (CEMS), Saitama 351-0198
}

\abst{
We theoretically study finite temperature properties of interacting fermion systems under geometrical frustration in the charge degree of freedom.
Physical quantities such as charge structure factors, the specific heat, and the entropy, of the two-dimensional model of interacting spinless fermions on an anisotropic triangular lattice are numerically calculated using the thermal pure quantum state. 
By considering the Coulomb interactions up to the next-nearest-neighbor bonds, we elucidate that in the highly frustrated region where a long-period stripe-type charge order (CO) is the ground state, fluctuations of different stripe-type CO patterns become large at finite temperatures. 
When we further introduce $1/r$-type long-range Coulomb interactions, such fluctuations are further enhanced and we find a region where both the stripe- and non-stripe-type CO fluctuations are prominent.
}

\begin{document}

\maketitle
\section{Introduction}
Organic conductors provide a fruitful field to study phenomena owing to strong electron correlations,\cite{Lebed2008,Seo2004} such as metal-insulator transition, magnetic/charge ordering, and superconductivity. One of their features is that the intermolecular bonds are mainly formed by the van der Waals interaction, then the molecules tend to accumulate in a close-packed manner. Therefore, structures inherently subject to geometrical frustration are often realized, most typically the triangular-based lattices. In fact, quantum spin liquid behavior is found in several organic materials, all having the anisotropic triangular lattice structure for $S=1/2$ localized spins under the influence of spin frustration\cite{KanodaKato2011}. On the other hand, when there exists half-integer number of electrons per lattice site, geometrical frustration can also act on the charge degree of freedom, called as the charge frustration effect\cite{Anderson1956}.

Quasi-two-dimensional organic conductors $\theta$-(BEDT-TTF)$_2$MM'(SCN)$_4$\cite{HMori1998} are such typical materials where systems with a half carrier (hole) per site on a nearly isotropic triangular lattice are realized\cite{Merino2005}. 
Here, BEDT-TTF stands for the bis(ethylene-dithio)tetrathiafulvalene molecule, M (M') takes ions such as Cs, Rb, and Tl (Co and Zn), and we abbreviate them as $\theta$-MM' hereafter.

In $\theta$-RbZn, a sharp metal-insulator transition accompanying lattice distortions occurs at a temperature ($T$) $\sim 195$~K, below which the so-called horizontal-stripe-type charge order (CO) is stabilized\cite{Miyagawa2000, Watanabe2004}. 
However, when the cooling rate is fast, the first-order CO transition can be quenched and an exotic state called the charge-cluster glass appears\cite{Kagawa13, Sato1378, Nogami2010, Chiba04, Nad2006}. On the other hand, $\theta$-CsZn does not show the sharp CO transition even under a normal cooling rate, but resembles this glassy state\cite{Suzuki2005, Chiba2008, Nad2008}. There, it is revealed that two types of CO fluctuations, corresponding to the horizontal stripe and a longer periodicity, coexist below $T\sim120$~K\cite{Watanabe1999, Sawano2005, Chiba2008, Nogami2010, Hashimoto2014}. By observing the cooling rate dependence of charge vitrification, cluster formation of the latter type of long-period CO is suggested to be related to the glassy nature in both salts\cite{Sato2013}. It is argued that a supercooled liquid of CO is realized  by kinetically avoiding the ``solid", i.e., the horizontal CO state, which is missed in $\theta$-CsZn within the laboratory accessible cooling rate. 

The relation between the geometrical charge frustration and the charge-cluster-glass forming ability has been pointed out experimentally.\cite{Sato2014} 
From X-ray diffraction measurements, the systematic variation in the anisotropy of the Coulomb interaction by changing anions suggests that the geometrical frustration promotes the slow charge dynamics.
Furthermore, a compound having a monoclinic crystal structure close to the isotropic triangular lattice, $\theta_{\rm m}$-TlZn, also shows similar crystallization and vitrification phenomena, even though the ``solid" here corresponds to a different CO pattern of diagonal-stripe type\cite{Sasaki1381}. 

Theoretically, the charge frustration in $\theta$-MM' has been intensively studied based on the 1/4-filled extended Hubbard model with intersite Coulomb interactions on the anisotropic triangular lattice\cite{Seo2006}. 
The existence of geometrical frustration destabilizes CO insulating states that are stabilized in the non-frustrated case such as the checkerboard-type CO\cite{Merino2005}. 
In models with short range Coulomb interactions, 
the ground state in the strongly frustrated region is the non-stripe-type CO state, so-called 3-fold CO state, at large values of the Coulomb interactions\cite{TMori2003, Kaneko2006, Watanabe2006, PhysRevB.78.035113}. This is stabilized by the kinetic energy term, when it is added to the classical ground state on the isotropic triangular lattice with macroscopic degeneracy; quantum fluctuation lifts the degeneracy. 
On the other hand, the long-period CO states mentioned above, are stabilized when the intersite Coulomb interactions up to next-nearest-neighbor bonds are included.\cite{TMori2003,Naka2014}

The ground state properties of such models have been investigated using various numerical methods such as mean-field theories, variational Monte Carlo methods, and exact diagonalization\cite{Seo2006}.
In association with the slow charge dynamics, the importance of long range nature of the Coulomb interaction have been discussed\cite{PhysRevLett.115.025701, doi:10.1143/JPSJ.81.063003, Fratini2009}. 
However, finite-$T$ analyses of charge frustrated systems considering the quantum effects are limited\cite{CanoCortes2010, CanoCortes2011, Merino2013, Naka2015}.
Especially, finite-$T$ properties of the long-period CO, which are important for understanding the charge fluctuations and the glassy nature as suggested by the experiments, have not been studied yet.

\section{Formulation}
\subsection{Model}
In this paper, we investigate finite-$T$ properties of interacting spinless fermions on a two-dimensional triangular lattice as shown in Fig. \ref{fig:model}(a). 
The model Hamiltonian is 
\begin{equation}
{\mathcal H} = - \sum_{i, j} t_{ij} c_{i}^{\dag}c_{j}+\frac{1}{2}\sum_{i, j}V_{ij} n_i n_j,
\end{equation}
where $t_{ij}$ is the transfer integral between $i$-th and $j$-th sites, $c_{i}^{\dag} (c_{i})$ is the creation (annihilation) operator of a spinless fermion at the $i$-th site, and the number operator is $n_i = c_i^{\dag}c_i$.
$V_{ij}$ is the Coulomb interaction between $i$-th and $j$-th sites.
\subsection{Treatment of $1/r$-type Coulomb interactions using Ewald's method}
In the following, we consider the nearest-neighbor transfer integrals along $x$ and $y$ axes and set $t_p = 0.2 $ eV, referring to $\theta$-MM'.\cite{HMori1998}
Hereafter, we set $1$ eV as an energy unit. We consider a finite-size cluster of lattice size $L = 6 \times 6$ and adopt the anti-periodic (periodic) boundary condition along $x$ ($y$)-axis to make the system closed shell. 
The average charge density $\langle n \rangle =\frac{1}{L}\sum_{i}\langle n_{i} \rangle$ is $0.5$.

\begin{figure}
\centering
\includegraphics[width=1 \columnwidth]{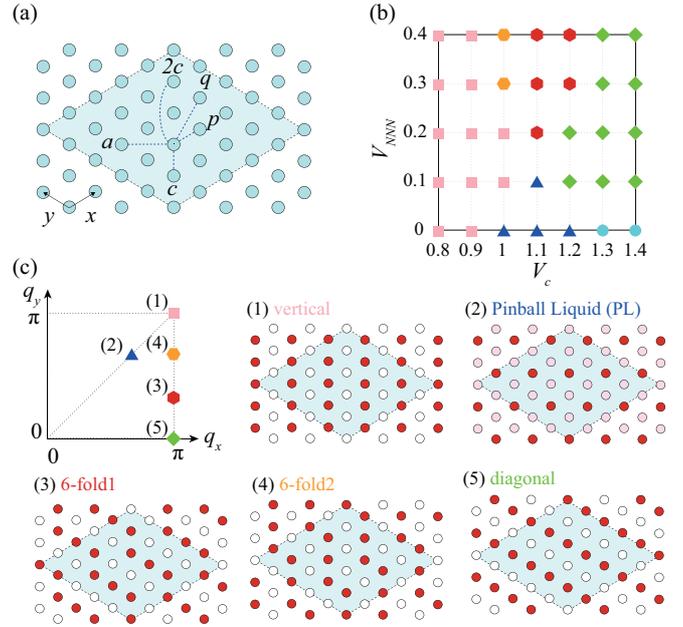}
\caption{(Color Online) (a) Schematic lattice structure of $\theta$-(BEDT-TTF)$_2$MM'(SCN)$_4$. Circles correspond to BEDT-TTF molecules. The shaded area shows the $6\times6$ cluster we treat in this work. Intermolecular bonds up to the next-nearest-neighbors are shown.  
(b) Ground state phase diagram on the $V_c$-$V_{NNN}$ plane for $V_p = 1$. The spatial charge order patterns defined in (c) stabilized in the ground state are indicated. 
(c) Top left figure shows the wave vectors of the CO states stabilized in our model, whose spatial patterns are drawn in Figures (1)-(5).The horizontal stripe state stabilized for $V_{NNN} =0, V_c \simeq 1.3$ are not shown here\cite{Naka2014}.}
\label{fig:model}
\end{figure}

In the following, we first consider the Coulomb interactions within the next-nearest-neighbor distance, i.e., on the bonds shown in Fig. \ref{fig:model}(a)\cite{Naka2014}. Then, we add the $1/r$-type long-range Coulomb interactions given as
\begin{equation}
V_{ij}(\lambda) =\sum_{\bm{R}} {}^{\prime} \frac{\lambda V_p}{|{\bm r_{ij} + {\bm R}}|},\label{eq:Ewald}
\end{equation}
where $V_p$ is the Coulomb interaction along the $p$ bonds, and $\lambda$ is a parameter to tune the strength of the Coulomb interactions.\cite{PhysRevLett.115.025701} ${\bm r}_{ij}$ and ${\bm R}$ indicate the relative vectors from $i$-th site to $j$-th site in the cluster and between different clusters, respectively.
Here the summation is taken over all clusters, but the Coulomb interactions within the next-nearest-neighbor distance are excluded.
By using the Ewald's method\cite{PhysRevLett.89.176803}, $V_{ij}$ is given by
\begin{align}
V_{ij} &= \left[\phi_{ij}^s+\phi_{ij}^L-\frac{2\sqrt{\pi}}{\alpha \Omega}-\frac{2\alpha}{\sqrt{\pi}}\delta_{i,j}\right] V_p, \label{Coulomb_eq}
\end{align}
where $\phi_{ij}^s =\sum_{{\bm R}} {}^{\prime} {\rm Erfc}( \alpha | {\bm r}_{ij} + {\bm R} | )/{| {\bm r}_{ij} + {\bm R}|}$, $\phi_{ij}^L =(2\pi/\Omega) \sum_{ \bm{\kappa} \neq {\bm 0}} 
{\rm Erfc}(|\bm{\kappa}/2\alpha|)\cos ({\bm \kappa} \cdot {\bm r}_{ij})/| {\bm \kappa}|$; $\Omega$ is the area of the cluster, $\alpha$ is the wave length for tuning the convergence of the series expansion of $\phi_{ij}^s$ and  $\phi_{ij}^L$ in Eq. (\ref{Coulomb_eq}), Erfc$(x)$ is the Gauss error function and $\bm{\kappa}$ is the reciprocal lattice vector of the cluster.
This setup of our model is based on the observation that the Coulomb interactions at short range sensitively reflect the anisotropy of the molecular orbitals, while they approach close to the $1/r$ form when the inter-molecular distance becomes far~\cite{TMori2003}.

\subsection{Numerical methods}

For the ground states, we perform the exact diagonalization using the Lanczos method. 
As for the finite-$T$ properties, we apply a recently developed method using the thermal pure quantum state\cite{PhysRevLett.108.240401}.
This method gives exact results in the thermodynamic limit but due to the finite-size effect, the results depend on the initial states. In this paper, we take $5$ random  initial states and estimate the finite-size effect by its standard deviations.

To identify the spatial pattern of CO at finite $T$,
we calculate the charge structure factors defined by
\begin{equation}
 N({\bm q},T) = \frac{1}{L}\sum_{i, j} \langle n_i n_j \rangle e^{-i {\bm q} \cdot {\bm r}_{ij}}.
\end{equation}
We also calculate the specific heat,
\begin{equation}
c(T)=\frac{\langle {\cal H}^2 \rangle-\langle {\cal H}\rangle^2}{L T^2},
\end{equation}
and the entropy,  
\begin{equation}
s(T)= \int_{0}^{T}\frac{c(T')}{T'}dT' = s(\infty)- \int_{T}^{\infty}\frac{c(T')}{T'}dT',
\end{equation}
where $s(\infty)$ is the entropy at the infinite-$T$ limit.
We define the normalized entropy as $\tilde{S}(T) \equiv s(T)/ s(\infty)$.

\section{Numerical Results}
\subsection{Effect of Coulomb Interactions up to the next-nearest-neighbor bonds}
We first show the results for the case up to the next-nearest-neighbor bonds. We set $V_p=1$, fix the ratio of Coulomb interactions as $V_{2c}:V_{q}:V_{a}=1.2:1:1$ and define $V_{NNN} = V_q = V_a$ [see Fig. \ref{fig:model}(a)]. These values  reproduce the long-period CO patterns close to those observed in $\theta$-MM'\cite{Naka2014}. 
\subsubsection{Ground state phase diagram}
Figure~\ref{fig:model}(b) shows the ground state phase diagram on the $V_c$-$V_{NNN}$ plane, which agrees with the previous result for a different cluster size\cite{Naka2014}. When $V_{NNN}=0$, in the frustrated region $V_c \simeq V_p$, a metallic CO state called the pinball liquid (PL) state is stabilized\cite{Hotta2006, doi:10.1143/JPSJ.75.123704}, whose wave vector is the same as in the 3-fold CO state; this state is suggested to show larger quantum fluctuation than the 3-fold CO state in the extended Hubbard model~\cite{CanoCortes2011}. This PL state is strongly suppressed by introducing $V_{NNN}$, while the long-period CO states, termed 6-fold1 and 6-fold2 CO states, appear at $V_c/V_p \sim 1$ where the degree of geometrical charge frustration is strong. 
The spatial patterns of these CO states are schematically shown in Fig.~\ref{fig:model}(c). Note that the patterns all have stripe-type CO with 1:1 ratio of rich and poor sites, except the PL with 1:2 ratio. The former are expected to result in insulating states in the strong coupling limit owing to the commensurability whereas the latter stays metallic.

\begin{figure}
\centering
\includegraphics[width=0.7 \columnwidth]{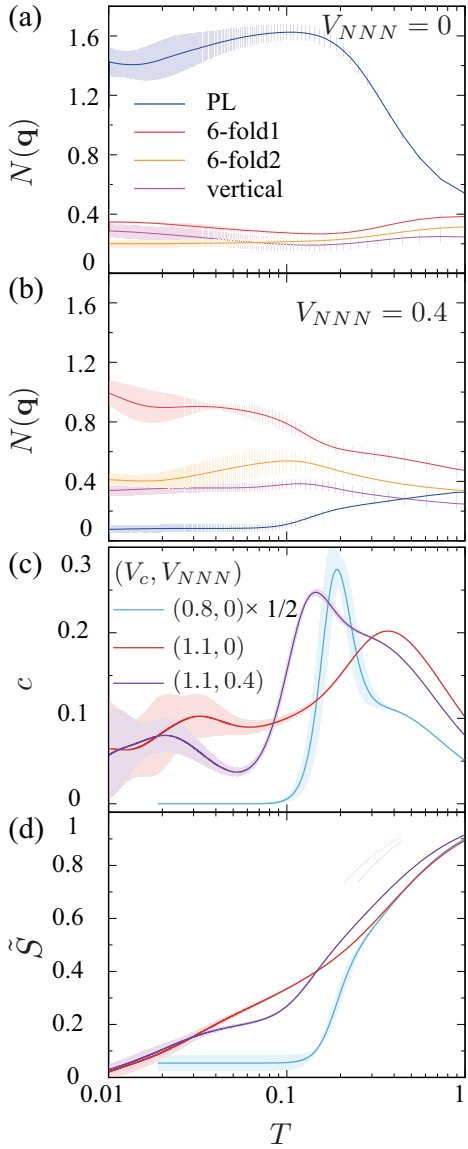}
\caption[width=1 \textwidth]{
(Color Online) $T$ dependencies of charge structure factors for different charge order states for (a) $(V_c, V_{NNN})=(1.1, 0)$ and (b) $(1.1, 0.4)$ for $V_p = 1$.
$T$ dependencies of (c) the specific heat and (d) the entropy for $(V_c, V_{NNN})=(0.8, 0), (1.1, 0),$ and $(1.1, 0.4)$ for $V_p = 1$. Error bars are shown as shaded areas\cite{errorbar}.}
\label{fig:2ndLRC}
\end{figure}

\subsubsection{Finite temperature properties}

Figures~\ref{fig:2ndLRC}(a) and \ref{fig:2ndLRC}(b) show the $T$ dependencies of charge structure factors $N({\bm q}, T)$ at $(V_c, V_{NNN})=(1.1, 0)$ and $(1.1, 0.4)$, where the 3-fold and the 6-fold1 CO are stabilized {in} the ground state, respectively.
In the former case, with decreasing $T$, the charge structure factor corresponding to the PL state is enhanced and other components become suppressed.
However, below $T \sim 0.15$,
the charge structure factor for the PL state turns to decrease, leading to a slight increase or saturation of other components at low $T$. This behavior is attributed to the so-called order by fluctuation mechanism\cite{Nagano2007}; the entropy gain due to a macroscopic number of low energy excited states stabilizes the order at finite $T$.
Such an entropy gain is expected in the PL state since carriers on its charge poor sites are mobile and therefore it is metallic.

For $V_{NNN}=0.4$ [Fig.~\ref{fig:2ndLRC}(b)], on the other hand, the 3-fold CO state is strongly suppressed, and the charge structure factors corresponding to the 6-fold1, 6-fold2, and vertical CO states gradually grow with decreasing $T$.
The development of the 6-fold1 CO structure factor shows a characteristic two-step behavior, at $T\sim0.15$ and $0.02$. 
This can be understood from the spatial CO patterns shown in Fig.~\ref{fig:model}(c). 
The 6-fold1 and 6-fold2 are both `tilted' stripe-type CO with close wave vectors, which have two kinds of charge rich sites with small and large charge transfer excitation energies [see Fig.\ref{fig:model}(c)]\cite{Naka2014}.
By raising $T$ from the ground state, first carriers on the sites with a smaller excitation energy fluctuate while the other charge rich sites remain stable, and at higher $T$ carriers on the latter sites also begin to fluctuate and the CO melts. This leads to the two-step behavior.

Such characteristic $T$ dependencies are manifested in the thermodynamic quantities as well. Figures~\ref{fig:2ndLRC}(c) and \ref{fig:2ndLRC}(d) show the specific heat and the entropy for $(V_c, V_{NNN})= (1.1, 0)$ and $(1.1, 0.4)$ as above, 
together with the case with weak frustration at  $(V_c, V_{NNN})=(0.8, 0)$ for comparison where the vertical-type CO state is stabilized as the ground state.
At the high $T$ region at $T \sim 1$, in all cases the specific heat gradually increases and the entropy is released with decreasing $T$. 
This indicates that the charge excitations with the energy loss at $V_p \sim 1$ are suppressed and the carriers begin to localize. 
This is consistent with the behavior of the charge structure factors, gradually growing as seen from Figs. \ref{fig:2ndLRC}(a) and \ref{fig:2ndLRC}(b).
As for the case of $(V_c, V_{NNN})=(0.8, 0)$, the specific heat has a sharp peak at $T \sim 0.2$, below which it rapidly approaches almost $0$ with decreasing $T$. The entropy also rapidly drops and becomes nearly $0$, indicating that the vertical CO state is well stabilized below $T\sim 0.2$ without much fluctuations.

In contrast, in the other two cases, the system does not show such a typical behavior seen for simple phase transitions and the effect of geometrical frustration is clearly demonstrated. For $(V_c, V_{NNN})=(1.1, 0)$, the specific heat has a broad peak at $T\sim 0.4$ where the charge structure factor of the PL state develops.
Below the peak, the specific heat does not decrease rapidly and the entropy is not released immediately.
The diffusive nature of the phase transition is attributable to the fluctuation in the PL state mentioned above.

Similarly, for $(V_c, V_{NNN})= (1.1, 0.4)$, the specific heat has a broad peak at $T \sim 0.15$ where  the 6-fold1 CO structure factor increases.
Below the peak, the specific heat decreases, apparently similar to the vertical CO case, but the entropy shows a slower decrease. 
However, the specific heat shows a further increase below $T \sim 0.05$ and the entropy remains as large as the case of the PL state in the low-$T$ limit. We consider that this low-$T$ behavior reflects the further development of the 6-fold1 CO seen in the structure factor, although numerical error becomes large in this region to be conclusive. 

\begin{figure}[t]
\centering
\includegraphics[width=0.8\columnwidth]{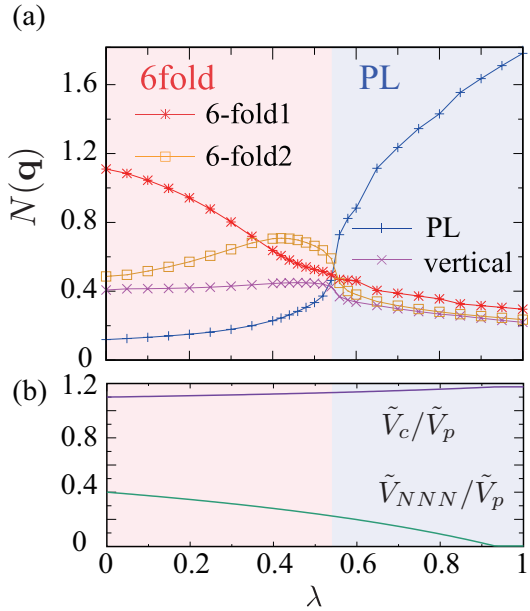}
\caption[width=1 \textwidth]{(Color Online) (a) $\lambda$ dependence of charge structure factors.
The Coulomb interactions within the next-nearest-neighbor distance are set as $(V_p, V_c, V_{NNN})=(1, 1.1, 1.4)$. 
(b) $\lambda$ dependence of the ratio of the effective Coulomb interactions within the Ewald's method. 
}
\label{fig:LRC}
\end{figure}

\subsection{Effect of $1/r$-type Coulomb Interactions}

Now, we investigate the effect of $1/r$-type long-range Coulomb interactions. Referring to the crystal structures of $\theta$-CsZn and $\theta$-RbZn \cite{HMori1998_2, Alemany2015}, we set the angle between $x$ and $y$ axis {as} $130$ degrees and estimate the Coulomb interactions by Eq.~(\ref{Coulomb_eq}). In the following, we fix $(V_p, V_c, V_{NNN}) = (1.0, 1.1, 0.4)$ and see how the $1/r$-type long-range Coulomb interactions work starting from the case where the ground state is the 6-fold1 CO state.

\subsubsection{Ground state phase diagram}
Figure \ref{fig:LRC}(a) shows the $\lambda$ dependence of charge structure factors at $T=0$. We can see that the ground state changes from the 6-fold1 to the 6-fold2 CO state, and then to the PL state with increasing $\lambda$.\cite{CommentGS}
In the region where two kinds of `tilted' CO states, i.e., the 6-fold1 and the 6-fold2 CO states, compete with each other, the charge structure factors of the vertical as well as the PL states become enhanced, suggesting a severe competition. 
However, above $\lambda\sim 0.55$, the ground state returns to the CO pattern when only nearest-neighbor Coulomb interactions are taken into account, e.g., at $(V_p, V_c, V_{NNN}) = (1.0, 1.1, 0)$. In fact, at $\lambda=1$, which corresponds to the fully long-ranged form, the charge structure factor for the PL state ($\sim1.8$) is even much larger than the case shown in Fig.~\ref{fig:2ndLRC}(a) at the ground state. This is a result which was unexpected considering a previous work on the role of long-range interactions in the classical limit, where stripe-type CO states are discussed to be stabilized.\cite{PhysRevLett.115.025701} We should note that, although the Ewald's method can treat long-range Coulomb interactions exactly treating the finite-size cluster, fluctuations larger than the cluster cannot be taken into account.

One of the origin of the recovery of the PL state with increasing $\lambda$ is the suppression of the effective next nearest Coulomb interactions. Since the Coulomb interactions larger than the size of the cluster are periodically added by using the Ewald's method, all pairs of Coulomb interactions in the cluster have finite values. As a result, the next nearest neighbor Coulomb interactions is effectively weakened compared to the nearest neighbor ones, and the PL state is restored.
To see this tendency, we show the $\lambda$-dependence of the ratio of the Coulomb interactions in. Fig.~\ref{fig:LRC} (b). Here we show the effective Coulomb parameters by $\tilde{V}_{ij}$, defined as those within the Ewald's method subtracted by the minimum value acting as a constant offset. With increasing $\lambda$, $\tilde{V}_c/\tilde{V}_p$ monotonically increases, while $\tilde{V}_a(\equiv \tilde{V}_{NNN})/\tilde{V}_p$ indeed decreases.

\subsubsection{Finite temperature properties}

\begin{figure}[ht]
\centering
\includegraphics[width=0.7\columnwidth]{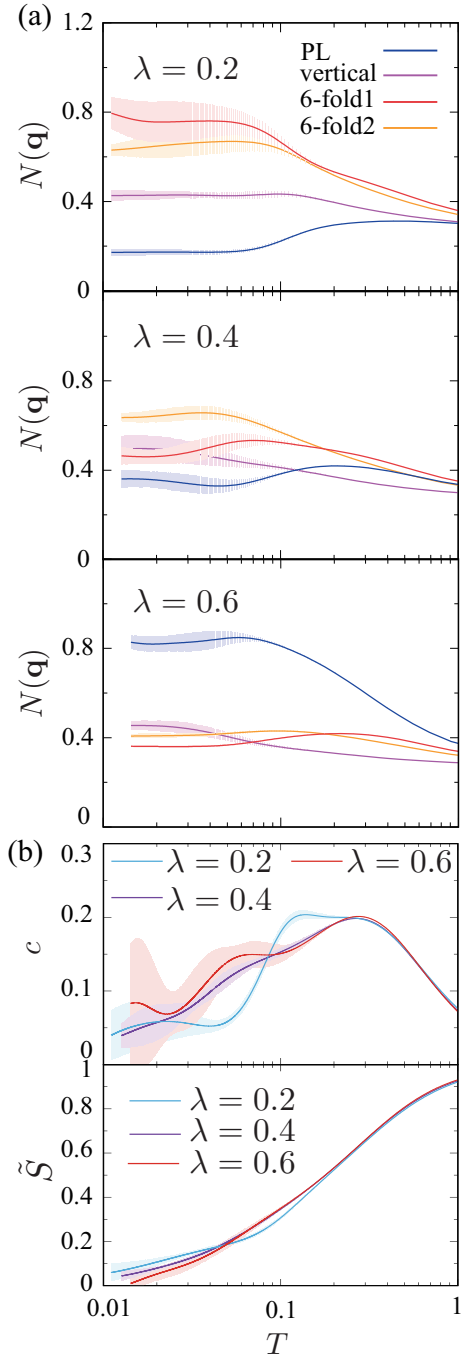}
\caption[width=1 \textwidth]{(Color Online) (a) $T$ dependencies of charge structure factors for different charge order states at $\lambda=0.2$ (top), $0.4$ (middle), and $0.6$ (bottom). 
The Coulomb interactions within the next-nearest-neighbor distance are set as $(V_p, V_c, V_{NNN})=(1, 1.1, 1.4)$. 
(b) $T$ dependencies of the specific heat (top) and the entropy (bottom) at $\lambda=0.2, 0.4, 0.6$.
In (a) and (b), error bars are shown as shaded areas\cite{errorbar}. 
}
\label{fig:LRC_finite_T}
\end{figure}

We show the $T$ dependencies of the charge structure factors in Fig. \ref{fig:LRC_finite_T}(a). First, by introducing $\lambda$, the charge structure factors of different kinds of CO patterns have almost the same values at high $T$ (e.g., $T>0.3$) compared to $\lambda=0$, i.e., the development of different kinds of CO patterns becomes featureless. With decreasing $T$ further, the charge structure factor for either 6-fold1 ($\lambda = 0.2$), the 6-fold2 ($\lambda = 0.4$), or the 3-fold ($\lambda=0.6$) CO state which is stable at the ground state becomes large. 
A noticeable point is that the fluctuation between different CO patterns become larger as we increase $\lambda$. For $\lambda=0.2$, the charge structure factors of the two 6-fold CO patterns becomes comparable. Moreover, when $\lambda$ is near the critical value such as in the data at $\lambda = 0.4$; the charge structure factors of CO with patterns not only the stripe-type but also for the PL state all show comparable values at finite $T$. For $\lambda=0.6$, on the other hand, the characteristic $T$ dependence of the PL state is seen, suggesting the order by fluctuation mechanism discussed above. 

Figure \ref{fig:LRC_finite_T}(b) shows the specific heat and the entropy. Reflecting the monotonic increase of charge structure factors with decreasing $T$ toward $T\sim 0.3$, the specific heat has a peak at $T \sim 0.3$ and the entropy is monotonically released. Below $T\sim 0.3$ for $\lambda=0.2$, the behavior is similar to the case for $\lambda=0$ and $(V_p, V_c, V_{NNN}) = (1.0, 1.1, 0.4)$ in Figs. \ref{fig:2ndLRC}(c) and \ref{fig:2ndLRC}(d), but with less pronounced feature in the two-peak structure, owing to the competition between two 6-fold CO states.
At $\lambda=0.4$, on the other hand, the specific heat remains large even at lower $T$, resulting in the residual entropy.
The $T$ dependence of the specific heat is nearly featureless, suggesting the maximized fluctuations between different CO patterns. By increasing $\lambda$ further, when the ground state is the PL state, the behavior resembles the case with only short range Coulomb interactions as the case of $(V_p, V_c, V_{NNN}) = (1.0, 1.1, 0)$ in Figs. \ref{fig:2ndLRC}(c) and \ref{fig:2ndLRC}(d).

\section{Summary and Discussion}
Finally, let us briefly discuss implications to experimental systems from our numerical results. 
As pointed out in refs.~\citen{TMori2003} and \citen{Naka2014}, the 6-fold CO states have close wave vectors to the long-period CO correlations experimentally inferred to be related to the charge-cluster glass formation in $\theta$-MM'. In our work, we have found that at finite $T$, characteristic behavior of such tilted stripe-type CO is seen as the $T$ driven fluctuations. When the model is limited to the next-nearest-neighbor Coulomb interactions, a two-step behavior of the development of charge structure factors is seen. This leads to a two-peak structure in the specific heat and results in a plateau-like behavior in the entropy. On the other hand, the addition of the $1/r$-type Coulomb interactions further enhance such finite-$T$ fluctuations, as long as it is below a critical value of $\lambda$, tuning its strength. We then speculate that the actual systems locate in such a region where different stripe-type CO states compete and fluctuations between them become large at finite $T$. The $1/r$ long-range Coulomb interactions indeed promote such fluctuation effect as proposed in previous studies.
However, its full inclusion in our setup recovers the PL state. We should note that there is a possibility that , in such a regime, fluctuations beyond  the cluster considered in this study becomes important.
Calculations treating larger cluster size to investigate it remains as a future problem.

\section{Acknowledgments}
\begin{acknowledgments}
	We thank T. Sasaki, K. Hashimoto, and T. Misawa for fruitful discussions. Numerical calculations were partly performed using the quantum lattice solver ${\mathcal H}\Phi$\cite{KAWAMURA2017180}, 
	at the Supercomputer Center, Institute for Solid State Physics, University of Tokyo.	
	KY was supported by Building of Consortia for the Development of Human Resources in Science and Technology, MEXT, Japan.
	This work was supported by JSPS KAKENHI Grant Numbers Nos. JP26400377, JP16H02393, and JP16K17731.
\end{acknowledgments}
\bibliography{ref}

\begin{thebibliography}{10}
\bibitem{errorbar}
Due to the finite-size effect, the calculation results obtained from thermal
  pure quantum states depend on the initial states. Thus, physical quantities
  are calculated by the mean values and the error bars of each quantity are
  also estimated by its standard deviations.
\bibitem{CommentGS}
For the anti-periodic boundary condition, the symmetry between $x$- and
  $y$-direction is broken. As a result, the charge structure factors show
  discontinuous changes at $\lambda \sim 0.3$ and $0.5$, which are artificial.
  To avoid this, for the ground state calculations, we show the results for the
  periodic boundary condition. On the other hand, finite temperatures
  calculations are performed using the anti-periodic boundary condition as in
  the rest of the paper.
\end{thebibliography}

\end{document}